\UseRawInputEncoding

\documentclass[conference]{IEEEtran}
\usepackage[T1]{fontenc}
\usepackage[utf8]{inputenc}
\IEEEoverridecommandlockouts
\usepackage{cite}
\usepackage{amsmath,amssymb,amsfonts}
\usepackage{algorithm}
\usepackage{algpseudocode}
\usepackage{graphicx}
\usepackage{multirow}
\usepackage{xcolor}
\usepackage{makecell}
\usepackage{enumitem}
\usepackage{placeins}
\usepackage[hidelinks]{hyperref}
\usepackage{dblfloatfix}
\usepackage{graphics}
\usepackage{balance}
\usepackage{booktabs}
\usepackage{longtable}
\usepackage{tikz}
\usepackage{array}
\usepackage{listings}
\usepackage{soul}
\usepackage{url}
\usetikzlibrary{shapes, arrows.meta, positioning}
\usepackage{xcolor}    
\usepackage{fancyvrb} 
\usepackage{mdframed} 
\usetikzlibrary{shapes.geometric, arrows}

\def\BibTeX{{\rm B\kern-.05em{\sc i\kern-.025em b}\kern-.08em
    T\kern-.1667em\lower.7ex\hbox{E}\kern-.125emX}}

\tikzstyle{startstop} = [rectangle, rounded corners, minimum width=3cm, minimum height=1cm, text centered, draw=black, fill=red!30]
\tikzstyle{process} = [rectangle, minimum width=3cm, minimum height=1cm, text centered, draw=black, fill=blue!30]
\tikzstyle{decision} = [diamond, minimum width=3cm, minimum height=1cm, text centered, draw=black, fill=yellow!30]
\tikzstyle{arrow} = [thick,->,>=stealth]

\definecolor{highlightblue}{RGB}{173, 216, 230}
\sethlcolor{yellow}    
\definecolor{lightblue}{RGB}{173, 216, 230}

\definecolor{codebg}{rgb}{1.0, 1.0, 0.85} 

\definecolor{keywordcolor}{RGB}{0, 0, 180}
\definecolor{stringcolor}{RGB}{163, 21, 21}

\definecolor{sectionColor}{RGB}{204, 229, 255}
\definecolor{expressionColor}{RGB}{255, 239, 213}
\definecolor{referenceColor}{RGB}{255, 220, 220}
\definecolor{statementColor}{RGB}{224, 255, 224}
\definecolor{ruleColor}{RGB}{255, 250, 205}
\definecolor{definitionColor}{RGB}{230, 230, 250}
\definecolor{exemptionColor}{RGB}{255, 228, 181}
\definecolor{infoColor}{RGB}{204, 255, 229}

\begin{document}

\title{Legal Requirements Translation from Law}


\author{
    \IEEEauthorblockN{Anmol Singhal, Travis Breaux}
    \IEEEauthorblockA{Carnegie Mellon University, Pittsburgh, USA\\ Email: \{singhal2, tdbreaux\}@andrew.cmu.edu}
}

\maketitle

\begin{abstract}
Software systems must comply with legal regulations, which is a resource-intensive task, particularly for small organizations and startups lacking dedicated legal expertise. Extracting metadata from regulations to elicit legal requirements for software is a critical step to ensure compliance. However, it is a cumbersome task due to the length and complex nature of legal text. Although prior work has pursued automated methods for extracting structural and semantic metadata from legal text, key limitations remain: they do not consider the interplay and interrelationships among attributes associated with these metadata types, and they rely on manual labeling or heuristic-driven machine learning, which does not generalize well to new documents.

In this paper, we introduce an approach based on textual entailment and in-context learning for automatically generating a canonical representation of legal text—encodable and executable as Python code. Our representation is instantiated from a manually designed Python class structure that serves as a domain-specific metamodel, capturing both structural and semantic legal metadata and their interrelationships. This design choice reduces the need for large, manually labeled datasets and enhances applicability to unseen legislation. We evaluate our approach on 13 U.S. state data breach notification laws, demonstrating that our generated representations pass approximately 89.4\% of test cases and achieve a precision and recall of 82.2 and 88.7 respectively. 

\end{abstract}

\vspace{-2mm}
\section{Introduction}

Technology companies develop products and services for users across multiple jurisdictions, each governed by distinct legal and regulatory frameworks. Ensuring compliance with diverse legal requirements is critical to avoiding sanctions and other penalties and ensuring uninterrupted operations. However, compliance verification is a resource-intensive process, particularly for small enterprises and startups lacking dedicated, in-house legal expertise. At the foundation of this challenge lies the need to accurately interpret and analyze legal text. Within requirements engineering (RE), structural and semantic metadata extraction from legal text plays an important role in systematically identifying a company's obligations, supporting traceability, and enabling the transition from regulatory documents to formal models, specifications, and eventually source code and other software artifacts. Structural metadata captures the hierarchical organization of legal provisions, which are often laden with nested statements, intricately inferred dependencies, and cross-references \cite{BG13}. On the other hand, semantic metadata encapsulates fine-grained attributes such as the entities regulated, deontic modalities assigned to actions, and pre- and post-conditions on actions \cite{SSS+21}. 

While prior work has explored taxonomies of individual metadata types and approaches to extract structural and semantic metadata attributes in isolation~\cite{SSS+21}, a major limitation has been the failure to maintain the contextual relationships between extracted elements, potentially leading to misinterpretation of regulatory obligations. For example, the data breach notification statute in the US State of Maryland (14–3504) consists of multiple layers of sub-rules and interdependencies (see Figure \ref{figur1:legal_excerpt}) that are necessary for a comprehensive understanding of legal text. Critically, legal meaning often emerges not from individual metadata elements alone, but from the relationships between them—such as how obligations, conditions, and exceptions interact across clauses and sub-sections~\cite{OA07, GB13}. Furthermore, existing approaches have largely relied on manual annotation or heuristic-driven machine learning (ML), both of which show weak generalization to unseen regulations. These challenges highlight the need for a structured and scalable approach to metadata extraction that not only captures the fine-grained semantics of legal text but also preserves its inherent structure and interdependencies for accurate and context-aware interpretation.

Recent advances in Large Language Models (LLMs) have demonstrated remarkable capabilities across various domains~\cite{Bro20}. These models can perform significantly well on complex tasks at inference time using in-context learning, eliminating the need for large labeled datasets for training~\cite{Bro20}. Despite emergent properties like Chain-of-Thought~\cite{KGR+22}, LLMs struggle with long-range dependencies~\cite{LLH+24,RSC+24} and logical consistency~\cite{JL23}, including those affecting implications~\cite{JQW+22}, negation~\cite{GHA+25} and transitivity~\cite{MNL+22}, making them unreliable for tasks requiring structured and factual outputs, such as legal compliance. To mitigate these issues, researchers have explored structured prompting techniques that encourage LLMs to generate outputs in predefined formats, such as JavaScript Serializable Object Notation (JSON), Python, or C++, instead of natural language. Prior work suggests that prompting LLMs to generate structured representations can help reduce hallucinations and generate reliable confidence estimates~\cite{KRM+24}. Moreover, these representations can enforce a defined schema, which eases parsing the output and limits the model's ability to generate irrelevant information. 

In this work, we propose an approach based on LLM code generation to generate a canonical representation of legal text automatically, which can be encoded and executed as a Python program. Our representation is designed to be minimal in complexity yet expressive enough to simultaneously capture the structural and semantic metadata of legal text. The approach consists of three key steps: (1) an iterative method to design a Python class structure for legal metadata representation, (2) a demonstration selection strategy based on textual entailment and cosine similarity to retrieve relevant exemplars for in-context learning, and (3) a code generation prompting technique to generate the encoded representation from a given legal provision. To the best of our knowledge, using a code-based representation to extract metadata elements and represent legal text has not been previously explored. 

We evaluate our approach on data breach notification laws from 13 US states. We extracted legal provisions from each law and manually annotated them with our representation to create a benchmark dataset, against which we assess the correctness of our approach. We propose an evaluation method incorporating unit testing and define 21 distinct test cases per sampled provision, each corresponding to a metadata attribute in our representation. We conduct experiments in two settings: (1) k-fold cross-validation on a development set of six state regulations and (2) testing on seven unseen state regulations. The results show that our approach passes 89.4\% of the total tests and significantly outperforms baselines. 

The main contributions of the paper are as follows:
\begin{itemize}
\item We propose a structured, executable representation to capture structural and semantic metadata, and preserve their interrelationships. 
    \item We propose a method based on textual entailment and in-context learning to automatically generate the representation of legal text.
    \item We propose an evaluation method based on unit testing to assess the correctness of the generated representation. 
    \item We analyze the trade-offs between structured code-based representations and natural language representations for legal metadata extraction.
\end{itemize}

The remainder of the paper is organized as follows: we present background and related work in Section~\ref{section:related_work}; we present our method and approach in Section~\ref{section:method} and our experimental setup in Section~\ref{section:evaluation}; we present results in Section~\ref{section:results} with discussion in Section~\ref{section:discussion}, followed by threats to validity in Section~\ref{section:threats} and conclusion in Section~\ref{section:conclusion}.

\begin{figure}[t]
\centering
\begin{mdframed}[linewidth=1pt]
\textbf{14–3504.} \\
\textbf{(d)(1)} The notification required under subsections (b) and (c) of this section may be delayed:
\begin{itemize}
    \item [(i)] If a law enforcement agency determines that the notification will impede a criminal investigation or jeopardize homeland or national security;
    \item [(ii)] To determine the scope of the breach of the security of a system, identify the individuals affected, or restore the integrity of the system.
\end{itemize}

\textbf{(2)} If notification is delayed under paragraph (1)(i) of this subsection, notification shall be given as soon as reasonably practicable after the law enforcement agency determines that it will not impede a criminal investigation and will not jeopardize homeland or national security.
\end{mdframed}
\caption{Maryland Personal Information Protection Act (\S 14–3504)}
\vspace{-3mm}
\label{figur1:legal_excerpt}
\end{figure}

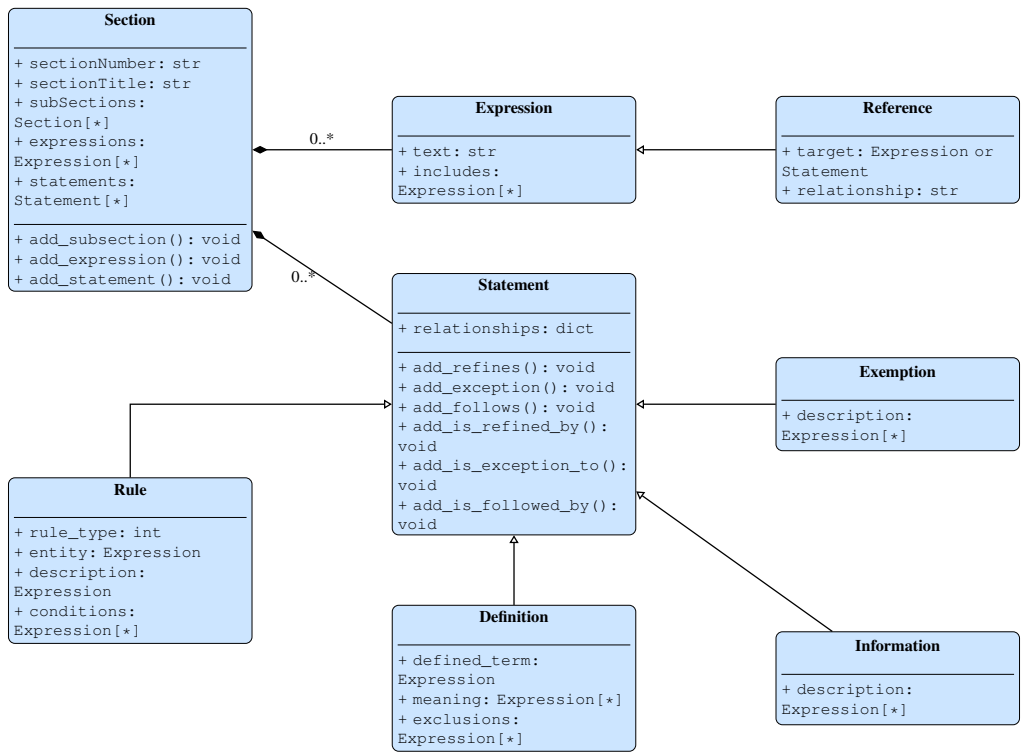
\begin{figure*}[ht!]
\centering
\resizebox{0.75\textwidth}{!}{
\begin{tikzpicture}[
    class/.style={rectangle, draw, rounded corners, text centered, text width=5cm, minimum width=5cm, minimum height=2cm, align=left},
    relationship/.style={->, thick},
    composition/.style={
    -{Diamond[fill=black,scale=1.2]},
    thick
},
inheritance/.style={
    -{Triangle[open]},
    thick
},
    node distance=3cm,
]

\node (Section) [class, fill=sectionColor] {
    \centering \textbf{Section} \\[0.5ex]
    \raggedright
    \rule{\linewidth}{0.4pt}
    \texttt{+ sectionNumber: str} \\
    \texttt{+ sectionTitle: str} \\
    \texttt{+ subSections: Section[*]} \\
    \texttt{+ expressions: Expression[*]} \\
    \texttt{+ statements: Statement[*]} \\
    \rule{\linewidth}{0.4pt}
    \texttt{+ add\_subsection(): void} \\
    \texttt{+ add\_expression(): void} \\
    \texttt{+ add\_statement(): void}
};

\node (Expression) [class, right=3cm of Section, fill=sectionColor] {
    \centering \textbf{Expression} \\[0.5ex]
    \raggedright
    \rule{\linewidth}{0.4pt}
    \texttt{+ text: str} \\
    \texttt{+ includes: Expression[*]} \\
};

\node (Reference) [class, right=3cm of Expression, fill=sectionColor] {
    \centering \textbf{Reference} \\[0.5ex]
    \raggedright
    \rule{\linewidth}{0.4pt}
    \texttt{+ target: Expression or Statement} \\
    \texttt{+ relationship: str}
};

\node (Statement) [class, below=1.5cm of Expression, fill=sectionColor] {
    \centering \textbf{Statement} \\[0.5ex]
    \raggedright
    \rule{\linewidth}{0.4pt}
    \texttt{+ relationships: dict} \\
    \rule{\linewidth}{0.4pt}
    \texttt{+ add\_refines(): void} \\
    \texttt{+ add\_exception(): void} \\
    \texttt{+ add\_follows(): void} \\
    \texttt{+ add\_is\_refined\_by(): void} \\
    \texttt{+ add\_is\_exception\_to(): void} \\
    \texttt{+ add\_is\_followed\_by(): void} \\
};

\node (Rule) [class, below=4 cm of Section, fill=sectionColor] {
    \centering \textbf{Rule} \\[0.5ex]
    \raggedright
    \rule{\linewidth}{0.4pt}
    \texttt{+ rule\_type: int} \\
    \texttt{+ entity: Expression} \\
    \texttt{+ description: Expression} \\
    \texttt{+ conditions: Expression[*]} \\
};

\node (Definition) [class, below=1.5 cm of Statement, fill=sectionColor] {
    \centering \textbf{Definition} \\[0.5ex]
    \raggedright
    \rule{\linewidth}{0.4pt}
    \texttt{+ defined\_term: Expression} \\
    \texttt{+ meaning: Expression[*]} \\
    \texttt{+ exclusions: Expression[*]}
};

\node (Information) [class, right=3cm of Definition, fill=sectionColor] {
    \centering \textbf{Information} \\[0.5ex]
    \raggedright
    \rule{\linewidth}{0.4pt}
    \texttt{+ description: Expression[*]}
};

\node (Exemption) [class, right=3cm of Statement, fill=sectionColor] {
    \centering \textbf{Exemption} \\[0.5ex]
    \raggedright
    \rule{\linewidth}{0.4pt}
    \texttt{+ description: Expression[*]}
};

\draw[composition, out=180, in=0] (Expression) to node[above] {0..*} (Section);
\draw[composition, left, midway] (Statement) to node[left] {0..*} (Section);

\draw[inheritance, out=180, in=0] (Reference) to node[above] {} (Expression);

\draw[inheritance] (Rule.north) |- node[above] {} (Statement.west);
\draw[inheritance, left, midway] (Information) to node[right] {} (Statement);
\draw[inheritance, out=90, in=270] (Definition) to node[left] {} (Statement);
\draw[inheritance, out=180, in=0] (Exemption) to node[above] {} (Statement);

\end{tikzpicture}

}
\caption{UML Class Diagram for Python Code Structure}
\label{fig:class-diagram}
\end{figure*}

\begin{figure*}[t]
\centering
\begin{minipage}{0.8\textwidth}
\begin{lstlisting}
s = Section(sectionNumber="14-3504.")
sd = Section(sectionNumber="(d)")
s.add_subsection(sd)
sd1 = Section(sectionNumber="(1)")
sd.add_subsection(sd1)
r1 = Rule(sd1, Expression(sd1, 
     "The notification required under subsections (b) and (c) of this section"))
r1.rule_type = Rule.PERMISSION
r1.description = Expression(sd1, "may be delayed")
ref1 = Reference(sd1, "subsection (b) and (c) of this section", target=r1)
r1.add_is_exception_to(ref1)

sd1i = Section("(i)")
sd1.add_subsection(sd1i)
r1.conditions.append(Expression(sd1i, 
     "a law enforcement agency determines that the notification will impede "
     "a criminal investigation or jeopardize homeland or national security"))

sd1ii = Section("(ii)")
sd1.add_subsection(sd1ii)
r1.conditions.append(Expression(sd1ii, 
     "To determine the scope of the breach of the security of a system, "
     "identify the individuals affected, or restore the integrity of the system"))

sd2 = Section(sectionNumber="(2)")
sd.add_subsection(sd2)
r2 = Rule(sd2, Expression(sd2, "notification"))
r2.conditions.append(Expression(sd2, 
     "notification is delayed under paragraph (1)(i) of this subsection"))
r2.rule_type = Rule.OBLIGATION
r2.description = Expression(sd2, 
     "shall be given as soon as reasonably practicable after the "
     "law enforcement agency determines that it will not impede a criminal "
     "investigation and will not jeopardize homeland or national security")
ref2 = Reference(sd2, "paragraph (1)(i) of this subsection", target=r2)
r2.add_follows(ref2)
\end{lstlisting}
\end{minipage}
\caption{Structured Representation of Section 14–3504 in Code Form}
\label{figure3:code_rep}
\vspace{-3mm}
\end{figure*}

\section{Background and Related Work}
\label{section:related_work}

\subsection{Legal Metadata}

Regulations and legal provisions dictate obligations, permissions, prohibitions, and constraints that must be adhered to, but their complexity makes manual compliance verification costly and error-prone. Automating compliance verification requires a structured understanding of legal provisions, which is where legal metadata extraction plays a crucial role.

Legal metadata refers to structured information that aids in the interpretation of legal provisions \cite{KL03}. Within RE, structural and semantic metadata are essential for systematically identifying obligations, supporting traceability, and enabling the transition from regulatory documents to formal specifications. 

\begin{itemize}
    \item \textbf{Structural metadata} captures the hierarchical organization of legal provisions, including sections, sub-sections, cross-references, and dependencies between rules.
    \item \textbf{Semantic metadata} encapsulates key legal elements such as deontic modalities (obligations, permissions, prohibitions), involved entities, conditions, and references to external provisions.
\end{itemize}

\subsection{Automating Legal Metadata Extraction}

\subsubsection{Structural metadata}

 Structural metadata is used mainly for establishing traceability to legal provisions ~\cite{BG13}, and for performing tasks such as requirements change impact analysis~\cite{SAS+17} and prioritization~\cite{M12, MOH+10}. Early methods for structural metadata extraction predominantly relied on rule-based systems that encode patterns to identify structural elements in legal texts. Akoma Ntoso~\cite{PV11} is a framework for representing structural metadata in legal texts that defines XML-based tags and mark structural elements such as $<$section$>$, $<$article$>$, $<$clause$>$, and $<$heading$>$. GaiusT~\cite{ZKM+15} employed pattern-based heuristics to identify section markers, references, and clause boundaries in regulatory texts. Recently, ML-based approaches have been proposed for structural metadata analysis. Chalkidis et al.~\cite{CFM+19} introduced the EURLEX57K dataset to facilitate hierarchical structure identification in European Union legislation.

\subsubsection{Semantic metadata}

Foundational work in requirements engineering relies on manual or semi-automatic methods for semantic metadata extraction. Breaux et al.~\cite{BVA06} tackle the elicitation of rights and permissions following the principles of deontic logic. Massey et al.~\cite{MOH+10, M12} developed an approach to map the terminology of a legal text onto that of a requirements specification. With the rise of ML and NLP in the last decade, research in legal text processing has increasingly applied ML and NLP pipelines to regulations. Bhatia et al.\cite{BEW+16} apply constituency and dependency parsing for analyzing privacy policies. Humphreys et al.~\cite{HBT+20} describe using Semantic Role Labeling to extract who-does-what information from legal texts and fill a legal ontology.  Sleimi et al.~\cite{SSS+21} proposed a harmonized view of semantic metadata in RE and used a hybrid approach involving NLP and heuristic-based ML to automatically extract semantic metadata from legal text. Amaral et al.\cite{CAA+22} developed an NLP-based approach that uses semantic frames to check a Data Processing Agreement (DPA) against GDPR obligations.

\subsubsection{Limitations of Existing Work}
While the proposed methods in prior work have contributed significantly to the field, they suffer from key limitations:
\begin{itemize}
    \item \textit{Losing Relationship between Structural and Semantic Metadata}: Most existing techniques extract metadata elements as independent fragments, often failing to maintain the hierarchical and logical relationships within legal provisions. For example, in Maryland’s Data Breach Notification Law (14–3504), different rules are interdependent. Extracting a single rule present in section (d)(2) in Figure \ref{figur1:legal_excerpt} without preserving its link to the preceding statement in section (d)(1) can lead to misinterpretations.
    \item \textit{Scalability Challenges}: Heuristic-based methods require extensive manual effort in designing extraction patterns and do not generalize well to new or unseen regulations. ML-based approaches often require large annotated datasets, which are expensive to create and maintain.
    \item \textit{Inability to Generate Actionable Representations}: Extracting metadata alone is not enough for compliance automation. There is a need for representations that encapsulate metadata in a structured format, allowing it to be used in downstream compliance verification tasks.
\end{itemize}

\subsection{Code-based Representation}

Given the challenges of legal text complexity, scalability, and reliability, our approach is based on textual entailment to generate a canonical, executable representation of legal provisions as Python code. Specifically, the generated code instantiates a manually constructed Python class hierarchy—i.e., a domain metamodel—that encodes structural and semantic metadata attributes relevant to legal requirements. We leverage in-context learning, a technique that uses examples within a prompt to train LLMs how to perform tasks at inference time. 

In recent years, several prompting techniques have been explored to use general LLMs for producing high-quality code, including zero-shot, few-shot, Chain-of-Thought (CoT), and self reflection~\cite{JWK+24}. LLMs pre-trained on code (Code-LLMs), such as OpenAI’s CodeX~\cite{CTJ+21} and DeepMind’s AlphaCode~\cite{LCC+22}, enable software engineers to automatically generate functions and sub-routines. A key benefit of using code LLMs is their ability to express natural language constraints for the desired functionalities that engineers want in their generated code~\cite{AON+21, JTM+22}. In addition to these natural language prompts, existing work has looked at directly providing code snippets as input to the model with an accompanying declarative instruction~\cite{JTW+23}. A recent prompting trend for code generation is to explicitly separate the problem-solving plan from the coding. Li et al.~\cite{LLL+25} proposed a method named structured CoT, based on using structured reasoning steps before outputting the final code. Another approach, called PAL~\cite{GMZ+23} (Program-Aided Language models), has the model output a piece of code (e.g. in Python) that, when executed, produces the answer to a problem. The success of PAL shows that asking the model to generate code as an intermediate step to solve a task can be useful for tasks that require deterministic model outputs. 

Our approach extends this idea of structured code representations to legal text, which, while more formalized than natural language, presents unique challenges such as nested dependencies, deontic modalities, and inter-paragraph references. In this work, we study the effectiveness of code generation to represent the relationships between structural and semantic metadata in legal text. Our approach offers several advantages:

\begin{itemize}
    \item \textit{Preserves Structural and Semantic Metadata}: By encoding legal text as Python code, we retain the hierarchical organization and dependencies between legal rules.
    \item \textit{Enhances Generalization Across Regulations}: Unlike heuristic-based methods, prompting LLMs to generate code dynamically adapts to different legal texts, improving generalizability to new jurisdictions and unseen laws.
    \item \textit{Actionable Representation of Legal Requirements}: Our representation can be executed by a Python interpreter, making it flexible to build custom visualizations of a legal provision for analysis, including representing relevant information as a knowledge graph.   
\end{itemize}

Importantly, we do not treat code translation as a method for semantic paraphrasing of the entire legal document. Instead, we define a canonical intermediate representation—formally specified via Python classes—that captures specific, pre-defined legal constructs (e.g., rules, references, conditions). This design parallels how code generation tasks restrict output to a constrained syntax to improve reliability \cite{GMZ+23}. 

\section{Method and Approach}
\label{section:method}

Our research method consists of three key steps: (1) discovering a Python class structure that encodes legal metadata; (2) developing a demonstration selection strategy using textual entailment and cosine similarity to retrieve relevant exemplars for in-context learning; and (3) designing a code generation prompt to generate structured legal encodings. The following sections detail the dataset creation and each step in the method.

\subsection{Legal Rule Code Corpus}
\label{dataset}
We selected 13 data breach notification laws from various U.S. states that govern the protection of residents' personal information. Our selection process followed the methodology described in \cite{BG13}, with the exception that we replaced the Alaska (AK) and Massachusetts (MA-S) laws included in their set with the California (CA) and Virginia (VA) laws. This change reflects the significant privacy law amendments enacted in California and Virginia over the past decade.

This dataset provides breadth in how laws are written while controlling for key criteria: every law is in English, follows the same legal system (Common Law), and addresses the same societal problem — under what conditions companies must notify data subjects about a breach of their personal information. While these regulations address the same theme, they vary in length, organization, and how they define legal entities and conditions for permitted, required, and prohibited actions. In distributed systems spanning these jurisdictions, such variations require businesses to decide how to comply. The selected laws are:

\begin{itemize}
\setlength\itemsep{2pt}  
\setlength\parsep{0pt}   
\small
    \item \textbf{Arkansas (AR):} Personal Information Protection Act, Arkansas Code §§ 4-110-101 et seq., enacted 2005.
    \item \textbf{Connecticut (CT):} CT: Breach of Security Regarding Computerized Data Containing Personal Information, Connecticut General Statute 36a-701b, enacted 2006.
    \item \textbf{Massachusetts (MA):} Security Breach Law, Massachusetts General Laws Chapter 93H, enacted 2007.
    \item \textbf{Maryland (MD):} Maryland Personal Information Protection Act, §§ 14-3501 et seq., enacted 2008.
    \item \textbf{Mississippi (MS):} Mississippi Consumer Data Privacy Act, Mississippi Code Annotated, §§ 75-24-29, enacted 2011.
    \item \textbf{Nevada (NV):} Security of Personal Information Law, Nevada Revised Statutes Chapter 603A, enacted 2006.
    \item \textbf{New York (NY):} Information Security Breach and Notification Act, New York General Business Law § 899-aa, enacted 2005.
    \item \textbf{Oregon (OR):} Oregon Consumer Identity Theft Protection Act, Oregon Revised Statutes §§ 646A.600–628, enacted 2007.
    \item \textbf{Utah (UT):} Protection of Personal Information Act, Utah Code §§ 13-44-101 et seq., enacted 2006.
    \item \textbf{Wisconsin (WI):} Notice of Unauthorized Acquisition of Personal Information, Wisconsin Statutes § 134.98, enacted 2006.
    \item \textbf{California (CA):} California Data Breach Notification Law, California Civil Code § 1798.82, enacted 2002.
    \item \textbf{Virginia (VA):} Breach of Personal Information Notification Act, Virginia Code § 18.2-186.6, enacted 2008.
    \item \textbf{Vermont (VT):} Security Breach Notice Act, Vermont Statutes Annotated, Title 9, § 2435, enacted 2006.
\end{itemize}

To construct our dataset, we used regular expressions to extract paragraphs from each legal document. Each paragraph can entail one or more legal statements that are related to one another and represent a legal requirement. We manually corrected any inconsistencies resulting from special characters, inconsistent punctuation, and other lexical errors. After the cleaning process, we obtained a total of 332 legal paragraphs.

\subsection{Creating the Python Class Structure}
\label{class structure}

We conceptualize the legal translation task as generating an instance of a formal metamodel tailored for regulatory texts. This metamodel is realized as a Python class structure designed to encode essential metadata types (e.g., rules, definitions, conditions) and their interrelationships (e.g., exceptions, refinements, follow-ups). Each generated legal translation is thus an instantiation of this metamodel, where class objects and their attributes capture the content of legal paragraphs.

The first author applied open coding~\cite{Sal12} to the corpus of legal paragraphs described in Section~\ref{dataset} to identify metadata attributes, yielding a coding frame consisting of 17 labels and corresponding definitions shown in Table~\ref{table:coding_frame}. This analysis began by assigning labels from existing taxonomies~\cite{SAS+17, BG13} to phrases in the legal text. While these taxonomies define several fine-grained semantic metadata attributes, we focused on those that occurred most frequently in our dataset. For example, in the excerpt depicted in Figure \ref{figur1:legal_excerpt}, the author identified the following labels: section, subsection, obligation, continuation, condition, reference, entity, and description. The coding frame was updated whenever the author encountered a new phrase without a proper label or if an existing definition required modification to cover the new phrase. The coding process continued until saturation was reached, which occurred when the author coded 150 paragraphs corresponding to seven laws without identifying new labels in the remaining 182 paragraphs.


\begin{table*}[t!]
\centering
\renewcommand{\arraystretch}{1.2}
\caption{Coding Frame with Labels and Definitions}
\label{table:coding_frame}
\begin{tabular}{l p{14cm}}
\toprule
\textbf{Tag Name} & \textbf{Description} \\
\midrule
\#definition & A legal statement defining the meaning of concepts \\
\#exclusion & A phrase highlighting what is excluded from the definition of a term \\
\#exemption & A legal statement that exempts someone/something from a rule \\
\#obligation & A statement imposing mandatory action to be performed by an agent \\
\#permission & A statement indicating the possibility to perform an action without an obligation or a prohibition \\
\#prohibition & A statement forbidding an action to happen or take place \\
\#penalty & A statement indicating the punishment for not following a rule \\
\#information & A legal statement about something that is known or proved to be true \\
\#continuation & Denoting nested legal statements; assigned whenever a phrase contains a colon and is followed by a bullet list \\
\#condition & A phrase in a statement highlighting a constraint under which a rule applies \\
\#follows & Relation that connects a statement to references or other statements that precede (act as pre-conditions to) the statement \\
\#refines & Relation that connects a statement that provides more information about a reference or base statement to the reference or base statement \\
\#followed\_by & Relation that connects a statement to references or other statements that follow the statement \\
\#refined\_by & Relation that connects a base statement to a cross-reference or another statement that provides more information about the base statement \\
\#exception & Relation that connects a statement to references or other statements that are exceptions to the statement \\
\#exception\_to & Relation that connects a statement that acts as an exception to a reference or base statement with the reference or base statement \\
\#reference & When the text contains pointers, numbers, or names to other sections, paragraphs, or laws \\
\bottomrule
\end{tabular}
\vspace{-2mm}
\end{table*}

After coding the dataset, the authors analyzed each label and the relationships among labels to identify classes and attributes. For example, obligations, permissions, prohibitions, and penalties were deemed as different kinds of rules. In addition, it was observed that these varieties have relationships to similar components represented in the text, such as entities who are the subject of the required or permitted action and conditions that must be true before the entity is required, permitted, or prohibited to perform the action. These observations led to the introduction of a \texttt{Rule} class with class attributes to include the \texttt{rule\_type}, \texttt{entity} and a Python list of \texttt{conditions}. Additionally, during the open coding exercise, the authors observed that exemption statements may not directly refer to or address specific entities, unlike rules. Consequently, exemption statements were mapped to a class separate from the \texttt{Rule} class. 

The paragraph structure of legal text often includes sentences that begin in one section and finish in another, complicating the translation of legal rules into data structures while preserving fine-traceability of which phrases come from which sections. To address this, a distinction was introduced with the \texttt{Expressions} and \texttt{Statement} classes. The \texttt{Expression} class encodes text within a legal paragraph, the smallest textual unit. Conversely, the \texttt{Statement} class can span multiple legal paragraphs if those paragraphs are nested under a larger one (i.e., the term \textit{continuation} has been previously used for this arrangement). Finally, paragraphs and the outer sections that contain them are organized hierarchically using the \texttt{Section} class.

The final Python class structure is depicted in Figure \ref{fig:class-diagram}. Docstrings were included to explain each class attribute, method, and data type. Based on the class diagram, the encoded representation for the Maryland statute is illustrated in Figure \ref{figure3:code_rep}. It demonstrates how the labels and legal text are aligned to yield the Python code as a translation of the law. 

\subsection{Demonstration Selection}
\label{demo_selection}

In-context learning with LLMs improves with demonstrations or examples that are added to the input prompt~\cite{LBM+22,ZWF+21}.
To create a dataset of demonstrations, the first author manually translated the dataset of legal provisions into the corresponding Python class structure. The author executed each code translation using a Python interpreter to ensure the syntactic correctness of the ground truth code. Due to the manual translation effort, the author required two weeks to complete the process and yield 332 legal text translations. On average, this involved approximately five minutes per paragraph to write and verify the code.

We partitioned the labeled dataset into a \textit{development set} comprising of 150 paragraphs that were translated before reaching saturation and a \textit{test set} consisting of the remaining 182 paragraphs, corresponding to six laws. We utilized the development set to sample demonstrations for in-context learning. The test set was used to evaluate our method. We have made the development and test sets publicly available. 

LLMs can exhibit strong performance improvements with just one demonstration on specific applications. In our experience, dependencies in translating legal text through legal metadata into Python are highly contextual, i.e., the presence or absence of one metadatum could change whether a later metadatum is expected (e.g., when a conditional phrase signals a subsequent rule but not a definition). In contrast, a multinomial labeling exercise that outputs one of a few labels is easier to demonstrate by sampling demonstrations across the labeling space. To address this challenge, we first use a zero-shot prompt to assign labels to a legal text paragraph drawn from the test set that needs to be translated. In Figure~\ref{fig:prompt1}, we show the prompt, which includes a task explanation, a dictionary of possible labels with definitions, the required output format, and an instruction to avoid generating explanations.

\begin{figure}[ht]
\begin{tabular}{p{8.5cm}}
\fontsize{7}{5}{\texttt{Read the text and assign tags based on the definitions provided. Do not create your own tags. }} \\
\fontsize{7}{5}{\texttt{Only output the tags in the form of a python list.}} \\
\fontsize{7}{5}{\texttt{Do not include the assigned parts of the text in your response.}} \\
\fontsize{7}{5}{\texttt{Tag Definitions: Python Dictionary containing tags and their definitions}} \\
\fontsize{7}{5}{\texttt{Text: input}} \\
\end{tabular}
\caption{Prompt to Label Legal Text}
\label{fig:prompt1}
\end{figure}

After generating labels for the input text paragraph, we retrieve demonstrations from the development set with the highest score, assigning one point for each matching label. From this list, we select three demonstrations with the highest cosine similarity score to the input text paragraph using the OpenAI text-embedding-3-large model. This demonstration selection method is motivated by the observation that high similarity scores are often correlated with legal text paragraphs sharing common labels. For example, two paragraphs defining ``personal information'' using the keyword ``means'' will typically yield a high similarity score, and both will be labeled as \#definition. We repeat this process for each legal text paragraph in the test set.

\subsection{Legal Text Translation Task}

We used GPT-4o, a closed-source LLM by OpenAI, to translate the input legal text paragraph into Python code. The translation prompt (see Figure~\ref{fig:prompt2}) includes a minimal instruction, the Python class definitions described in Section~\ref{class structure}, and the demonstrations describing example translations from legal text into Python that were selected using the strategy described in Section~\ref{demo_selection}. The instruction explicitly discourages classes or class attributes not shown in the structure. Owing to the deterministic nature of the translation task, we set a temperature parameter 0.5 for our experiments.

\begin{figure}[ht]
\begin{tabular}{p{8.5cm}}
\fontsize{7}{5}{\texttt{Read the text and convert it to Python code. Use the class structure detailed below to write code. Do not create your own names. Examples have been provided. }} \\
\fontsize{7}{5}{\texttt{Class Structure: Python class structure enclosed in triple back-ticks}} \\
\fontsize{7}{5}{\texttt{Examples: the three sampled demonstrations}} \\
\fontsize{7}{5}{\texttt{Text: input}} \\
\end{tabular}
\caption{Prompt to Translate Legal Text into Python}
\label{fig:prompt2}
\vspace{-3mm}
\end{figure}

\section{Experimental Evaluation}
\label{section:evaluation}

We evaluate the approach by answering the following research questions (RQs):

\begin{itemize}
\item \textbf{RQ1}: How accurate is the generated Python code?
\item \textbf{RQ2}: How does the method compare to traditional LLM text extraction using JSON?
\item \textbf{RQ3}: To what extent do the method steps contribute to overall performance improvement?
\item \textbf{RQ4}: To what extent does the method generalize to unseen legal texts?
\item \textbf{RQ5}: What are the sources of error in the steps within the method?
\end{itemize}

We now describe the development of our evaluation strategy by first introducing our use of unit tests to score the generated Python code, followed by the metrics used to compute scores, before describing the two methods for evaluation.

\subsection{Unit Testing-based Evaluation}
\label{testing}

Translating legal text into code provides the benefit that the generated representation can be analyzed programmatically using traditional test harnesses. In our case, we adopt a unit testing framework to perform \emph{model conformance checks}, i.e., verifying whether the generated code correctly instantiates the expected legal metadata structure and semantics.

Although unit tests are typically associated with verifying execution behavior, we use them to validate that each generated code instance conforms to the manually defined Python class structure (described in Section~\ref{class structure}) and matches ground truth attribute values. 

Our evaluation proceeds in three steps. First, we manually author test cases for each attribute defined in the class structure. Then, we generate the Python code for a legal paragraph (see Figure~\ref{fig:prompt2}), execute it using the Python interpreter, and serialize the resulting instantiated objects. Finally, we run the test cases on the serialized objects to compare each generated attribute value with the corresponding value in the ground truth. Thus, each test validates a specific structural or semantic property of the instance.

We designed three categories of tests:

\begin{itemize}
\item \textbf{Compilation Test}: This test verifies whether the generated code is syntactically correct with respect to the predefined class structure. The code passes this test if it executes without errors in the Python interpreter. Failures typically arise from hallucinations, for example, references to undefined classes, methods, or attribute names.

\item \textbf{Structural Tests}: These tests check if each generated class has the expected minimal attributes (e.g., a \texttt{Definition} must contain a \texttt{term}, regardless of the value assigned). The list of attributes is shown in Table~\ref{table:coding_frame}. Structural tests are paragraph-independent and include five tests.

\item \textbf{Semantic Tests}: These tests verify whether the attribute values generated for each paragraph semantically match those of the ground truth. We used an exact match comparison after normalizing the strings (lowercase, stop-word removal, and punctuation stripping). For example, words such as ``means'' or ``if'' are removed before comparison. Semantic tests include 16 paragraph-independent checks. An example of a semantic unit test to identify the correct references in the legal excerpt in \ref{figur1:legal_excerpt} will check if the objects of class Reference are initialized with the same values in the generated and ground truth code (subsections (b) and (c), paragraph (1)(i) in this case).
\end{itemize}

Each generated paragraph is evaluated using this suite of 22 tests: one compilation test, five structural tests, and sixteen semantic tests. A paragraph-level output is considered fully correct only if all 22 tests pass. Although implemented using a unit testing framework, the purpose of these tests is to validate the conformance of the generated code to a well-defined and interpretable metamodel, rather than to test the functional execution or control flow behavior.


\subsection{Evaluation Metrics}
\label{metrics}

We computed the following metrics for each paragraph in our development and test sets:
\begin{itemize}
\item \textbf{Overall accuracy}: The number of tests passed divided by the total number of tests executed.
\item \textbf{Attribute Precision}: The number of semantic tests passed where the attribute in the generated code was present in the ground truth code divided by the total number of semantic tests where the attribute was \textit{present in the generated code}.
\item \textbf{Attribute Recall}: The number of semantic tests passed where the attribute in the generated code was present in the ground truth code divided by the total number of semantic tests where the attribute was \textit{present in the ground truth code}.
\end{itemize}

We report average scores at the paragraph level and attribute level in the results in Section~\ref{section:results}.

In addition to the above tests, we compute pass@k~\cite{CJ+21}, a standard metric used in the evaluation of generative code tasks. This metric is particularly relevant for LLM-based code generation, where sampling is non-deterministic, meaning that the same prompt can yield different outputs across multiple runs. Pass@k asks whether any of the $k$ sampled generations yields a completely correct code translation, i.e., one that passes all unit tests for that paragraph.

Pass@k is important to account for stochasticity in LLM outputs and better reflects real-world use cases where multiple samples can be taken to increase reliability. For example, even if a single generation is imperfect, users might run the model multiple times to obtain a valid solution. 

Precision and recall are computed for attributes that are present in both the generated and ground truth code. These metrics provide insight into the partial correctness of the outputs, which is useful for debugging specific failures and identifying attribute types that are more difficult to generate. However, they do not reflect whether an entire output is usable or not. Accuracy measures the proportion of tests passed across all test cases and paragraphs but similarly does not capture whether a complete, semantically accurate representation has been generated for each paragraph. Therefore, we report both standard metrics (precision, recall, accuracy) and pass@k score to give a more holistic picture of the correctness and practical usability of the method.

To measure pass@k, we test whether the generated code passes all the 21 syntactic and semantic test cases, i.e., that the model output correctly generates all the attributes present in a paragraph within $k$ prompt executions. The pass@k score is computed by generating $n$ code solutions per paragraph, where $n$ must be greater than or equal to $k$. Let $c$ be the number of solutions that pass all unit tests, where $c$ is less than or equal to $n$. The pass@k score is calculated using the following equation:

\begin{itemize}
    \item $n$ is the number of coding solutions;
    \item $c$ is the number of solutions that passed all unit tests out of $n$;
    \item $k$ is the number of $k$ combinations selected from $n$;
    \item $comb(n-c, k)$ is the number of $k$ combinations that fail to pass the unit test, chosen from $n$;
    \item The fraction $comb(n-c, k) / comb(n, k)$ represents the probability that all $k$ generated solutions fail the unit test.
    \vspace{-5mm}
\end{itemize}

\begin{equation}
   pass@k = E( 1 - ( comb(n-c, k) / comb(n, k) ) \\ 
\end{equation} 

We compute pass@k by varying the value of $k$ from one to five. We assumed that the values of $k$ and $n$ are equal, meaning that we sample $k$ model outputs per paragraph in the dataset. We report the paragraph-level metrics for the samples that passed the maximum number of test cases. 

\subsection{Evaluation Strategy}
\label{benchmarks}
The evaluation strategy consists of two evaluation methods:

\begin{itemize}
\item \textbf{Five-Fold Cross Validation on Development Set}: Because the translation of legal text to code can yield inconsistencies in the expected output that are sensitive to the prompt design and the class structure, we chose to fully employ the development set to refine the prompts and class structure. We divided the development set into five folds through random sampling, and for five iterations, four folds (amounting to 120 paragraphs in our dataset) served as the demonstration sampling dataset, while one held-out fold (consisting of 30 paragraphs) served as the test set to obtain model output. We report the average scores obtained across all five folds to answer RQ1.
\item \textbf{Document-Level Evaluation on Test Set}: We evaluate the documents in the test set individually and compute scores for each document separately to address RQ2. We report both the cumulative average scores obtained across documents in the test set.
\end{itemize}

We answer RQ1 by computing the metrics in Section \ref{metrics} for the development and test set. We report the results for RQ2, RQ3, and RQ4 on the test set. To the best of our knowledge, no prior work has evaluated code-generation models on the task of translating legal text into structured representations. As such, we cannot directly compare our results for RQ2 to those of existing techniques on this task. Therefore, to answer RQ2, we run a baseline prompt (see Figure~\ref{fig:prompt3}) in which we instruct the model to assign values to each attribute using a JSON schema. We provide the model with the definitions of each attribute and specify the input and output format. The model output is serialized into a dictionary and compared with the serialized ground truth code. 

\begin{figure}[ht]
\begin{tabular}{p{8.5cm}}
\fontsize{7}{5}{\texttt{Read the provided text and identify the portion of text that corresponds to each attribute explained below. If the text does not contain a specific attribute, ignore it and move to the next one. Include all the parameters for each attribute in your response. The output should be in the form of a JSON list. An example has been provided. }} \\
\fontsize{7}{5}{\texttt{Attribute Definitions: }} \\
\fontsize{7}{5}{\texttt{Attribute Parameters: }} \\
\fontsize{7}{5}{\texttt{Example: }} \\
\fontsize{7}{5}{\texttt{Text: input}} \\
\end{tabular}
\caption{Baseline Prompt for Direct Attribute Extraction}
\label{fig:prompt3}
\end{figure}

\begin{table*}[t!]
    \centering
    \caption{Evaluation Results for Different Approaches}
    \begin{tabular}{l c c c c c c c c}
        \toprule
        \textbf{Dataset} & \textbf{Approach} & \textbf{Compilation Test} & \textbf{Structural Test} & \multicolumn{3}{c}{\textbf{Semantic Test}} & \textbf{Pass@3} \\
        \cmidrule(lr){4-6}
        & & & \textbf{Accuracy} & \textbf{Precision} & \textbf{Recall} & \\
        \midrule
        Test & Text-gen (JSON baseline) & NA & 54.2\% & 72.1\% & 68.5\% & 61.0\% & 31.2\% \\
        Test & Code-gen + class & 98.5\% & 75.0\% & 83.7\% & 80.4\% & 67.4\% & 38.0\% \\
        Test & Code-gen + demo & 85.7\% & 82.8\% & 82.0\% & 73.8\% & 72.8\% & 42.1\% \\        
        Dev & Code-gen + class + demo & 98.4\% & \textbf{83.4\%} & 86.9\% & 79.1\% & 85.3\% & 56.7\% \\
        Test & \textbf{Code-gen + class + demo} & \textbf{99.2\%} & 82.0\% & \textbf{89.4\%} & \textbf{82.2\%} & \textbf{88.7\%} & \textbf{62.1\%} \\
        \bottomrule
    \end{tabular}
    \label{table:rq1_results}
\end{table*}

We answer RQ3 by conducting two ablation studies:
\begin{itemize}
    \item \textbf{Class Structure}: We compare our approach with a baseline in which we remove the class structure as context from the prompt and instructed the model to convert the text to code only on the basis of the demonstrations. 
    \item \textbf{Demonstration Selection}: We compare our approach to a baseline in which we replace the demonstration selection strategy with a keyword-based approach to select examples to prompt the LLM. The keywords represent frequently occurring phrases in the text that could trigger the assignment of a label. For example, `if' and `whether' are keywords for assigning the \#condition label. 
\end{itemize}

\section{Results}
\label{section:results}

Table~\ref{table:rq1_results} presents the overall accuracy, precision, recall and pass@$k=3$ results: the Approach column includes the \textit{Code-gen + class + demo} study, which is the end-to-end study providing the class structure and demonstration strategy to select demonstrations; the Text-gen study evaluates the traditional LLM-based text extraction of the attributes using a JSON schema; the \textit{Code-gen + demo} study is the first ablation study in which the the class structure is removed, and the \textit{Code-gen + class} study is the second ablation study in which the class structure is presented, but the demonstrations are randomly sampled, ignoring the selection strategy.

The RQ1 asks ``how accurate is the generated Python code structure?''. In Table~\ref{table:rq1_results}, we see that the attribute-level accuracy, precision, and recall are all highest for the complete method, and outperform the Text-gen approach by a margin of 30\% on the pass@k score. This result also answers RQ2, which asks ``How does the method compare to traditional LLM text extraction using JSON?''. The ablation results in Table~\ref{table:rq1_results} show that the end-to-end method benefits from all of its features, including the class structure and demonstration strategy, in response to RQ3 that asks ``To what extent do the method steps contribute to overall performance improvement?''. 

\begin{figure}[t]
     \centering
     \includegraphics[width=0.8\linewidth]{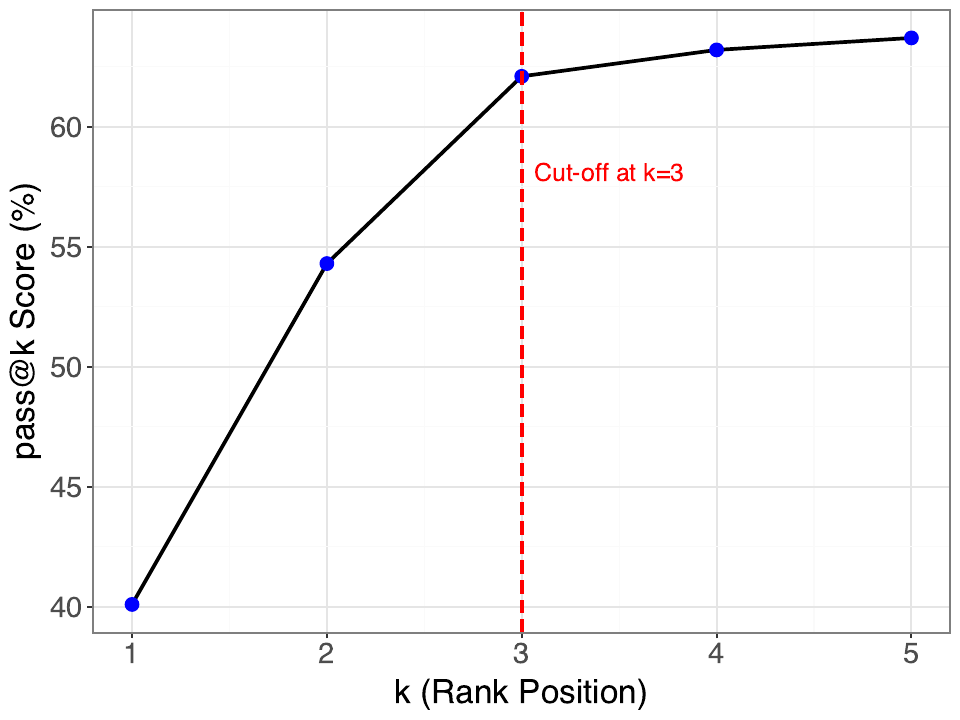}
     \caption{Pass@k score trend on the test set}
     \label{fig:graph}
     \vspace{-3mm}
 \end{figure}

The pass@k score trend, which denotes the number of paragraphs in the data set that passed all test cases in $k$ attempts is reported in Figure \ref{fig:graph}. For only one generation per paragraph, GPT-4o yields a pass@1 score of 40\%. The score improves significantly as we increase the value of $k$ to $k=3$, and only marginally improves for $k>3$.

For $k=3$, our approach demonstrates a high accuracy across compilation, structural, and semantic tests on both the development and test sets (Table \ref{table:rq1_results}). The compilation test reported a near-perfect score, which shows that GPT-4o reliably generates executable Python code based on the class structure. The structural test and semantic test accuracy of approximately 90\% on the total tests conducted shows that our approach generates a representation that closely resembles the ground truth representation. The recall is slightly lower than the precision. The error analysis is reported in Section~\ref{section:discussion}. 

The pass@3 score was approximately 56.7\% and 62.1\% on the development and test sets, respectively. These scores indicate that while the generated code matched the ground truth for most attributes, a few attributes demonstrated a higher failure rate, lowering the overall correctness of the model output. We discuss the implications of these findings in Section~\ref{section:discussion}.

\begin{center}
\begin{table}[ht]
\centering
 \renewcommand{\arraystretch}{1.2}
\caption{Results For Each Attribute}
\resizebox{0.9\columnwidth}{!}{%
\begin{tabular}{ l c c c } 
 \toprule
  \textbf{Attribute} & \textbf{Accuracy} & \textbf{Precision} & \textbf{Recall} \\ 
  \midrule
Information               & 96.9\% & 50.0\% & 25.0\% \\ 
Definition term            & 97.6\% & 100\% & 93.4\% \\ 
Definition meaning         & 95.2\% & 100\% & 92.8\% \\ 
Definition exclusions      & 97.6\% & 50.0\% & 50.0\% \\ 
Exemption                  & 85.9\% & 37.5\% & 37.5\% \\ 
Rule Entity                & 86.6\% & 85.3\% & 94.4\% \\ 
Rule Description           & 78.7\% & 78.8\% & 93.7\% \\ 
Rule Condition             & 78.7\% & 87.5\% & 83.3\% \\ 
Rule Type                  & 89.0\% & 86.1\% & 94.4\% \\ 
References                 & 74.8\% & 58.3\% & 100\% \\ 
Relationship type          & 85.0\% & 33.1\% & 45.8\% \\
No. of sections            & 93.0\% & 91.5\% & 92.5\% \\ 
No. of subsections         & 87.0\% & 85.0\% & 86.0\% \\ 
No. of statements          & 89.0\% & 87.5\% & 88.0\% \\ 
No. of expressions         & 84.5\% & 81.0\% & 82.5\% \\ 
Section naming             & 91.0\% & 88.0\% & 89.5\% \\ 
\bottomrule
\end{tabular}%
}
\label{table:class_res}
\end{table}
\vspace{-4mm}
\end{center}

RQ4 asks ``To what extent does the method generalize to unseen legal texts?'' Across six legal texts held out in the test set, we observe that the method performed better than the development set by +6\% on the pass@3 metric. The increase could be attributed to our design choice, where a higher number of legal paragraphs were used to sample demonstrations on the test set (150 paragraphs) than the validation set (120 paragraphs). We further discuss this finding in Section \ref{section:discussion}. 

We also compare the test set results for each attribute in the dataset in \ref{table:class_res}. We report the aggregate score for relationship type by averaging the individual scores for each of the six types of relationship we observed in the dataset. The method performed best across definitions and the number of section-related tests. The results were primarily poor across exemptions, references, rule conditions, and relationship types, which we explain in the discussion section. 

\section{Discussion}
\label{section:discussion}

Our findings demonstrate that a code-based representation of legal text can substantially improve both the structural and semantic accuracy of extracting legal metadata. Below, we discuss the implications of these results and highlight the potential benefits of this representation.

\subsection{Error Analysis}

We conducted a detailed error analysis of the errors obtained on the test set. We identified 20 unique errors that are broadly categorized into four types:
\begin{itemize}
    \item \textit{Reference-related errors}: Errors that include missing reference objects, incorrect relationship assignment, and exhibit detailed expressions as references. 
    \item \textit{Sentence decomposition inconsistency}: Failure to identify dependent clauses for entities and actions, such as conditions and exceptions, causing a high mismatch error rate across these attributes. 
    \item \textit{Inference-related errors}: Incorrect statement type, incorrect rule type, incorrect relationship between statements 
    \item \textit{Sectioning-related errors}: Flattening hierarchical structure and missing clauses
\end{itemize}

A root cause analysis shows that a possible cause of these errors is the imperfect retrieval of high-quality examples. While the demonstration selection strategy boosts the performance of GPT-4o on the code generation task, the earlier tagging task is less reliable and can produce incorrect labels, leading to the selection of the wrong demonstrations. To make the step more deterministic, prompting techniques, such as Chain-of-Thought~\cite{KGR+22}, or fine-tuning a separate classifier to predict labels may be useful. 

\subsection{Takeaways from Results}

Our conformance-based unit testing approach ensures that the generated representation is well-formed, complete, and semantically aligned with the source text, even though it does not simulate rule execution. This is analogous to unit tests in code synthesis pipelines where the goal is to verify correct API usage, structural properties, or data flow conformance before execution-level behaviors are modeled.

The high accuracy in structural and semantic tests shows that prompting GPT-4o to generate class-based code structures yields outputs closely mirroring the ground truth annotations. By encoding legal statements as Python objects, our approach enforces rigorous formatting and data-typing constraints in the narrower vocabulary of code. The near-perfect compilation rate suggests that the structured prompt and class definitions improve task selection at inference time, reducing free-form text errors like hallucinations or incomplete clauses that may occur in text-to-text generation.

While paragraph-level accuracy, precision, and recall metrics show our approach closely resembles the ground truth, the pass@k scores plateau near 62\% on the test set. This result may be due to the strict evaluation, which measures total correctness using an exact match on ground truth attribute values, meaning that all attributes must be identified exactly as represented in the ground truth code.

Although the approach achieved strong scores for most attributes, references and conditional clauses showed comparatively higher error rates. This highlights the complexity of legal language and cross-referencing. Unfortunately, these mechanisms communicate critical exceptions that trigger only under specific conditions. Developing targeted strategies for reference extraction may be needed to manage deeply nested and multi-layered cross-references. Combining textual entailment with self-reflection~\cite{ZYX+23, RG24} could potentially address some of these errors.

Our method shows a significant improvement over the JSON baseline. This finding is consistent with prior work, which shows that flattening structured representations into text tends to reduce task accuracy \cite{MZA+22}. This decline occurs because: (1) serialized structures are underrepresented in pre-training data compared to free-form text, and (2) flattening a structured graph often separates semantically related nodes, found close together in the graph, across distant positions in a flat string.

\subsection{Potential Benefits of a Code-Based Representation}

By translating legal provisions into Python code, our framework produces an ``executable'' representation of regulatory text. Because each generated output is an instantiation of a formally defined metamodel, our method offers a new avenue for compliance automation: legal requirements captured in code could be integrated with testing frameworks and model-driven engineering tools. Moreover, once encoded as Python objects, provisions are amenable to further transformations, such as generating visual dependency graphs, programmatically verifying contradictions, or enforcing traceability by tagging software artifacts.

In-context learning leverages a small set of carefully selected demonstrations, thus diminishing the need for large, manually labeled datasets. In settings where domain experts (e.g., legal counsel) are scarce, this approach lowers the overhead of preparing corpora by requiring attorneys to annotate texts. Rather, the generated code could be visually represented and inspected by attorneys to save time and focus their feedback on structural representations more amenable to software developers. It also allows for agile updates: when new regulations arise, only a few additional exemplars are needed for the model to adapt, rather than retraining an entire pipeline.

Legal text differs from general natural language in its use of formal constructs and well-defined logical dependencies among obligations, exceptions, and definitions. Our empirical results show that these properties make legal text particularly suitable for deterministic translation into structured representations. Representing these constructs through a unified class hierarchy (e.g., separate classes for rules, exemptions, references) ensures that the nested relationships remain explicit. Generating Python code that instantiates well-defined classes provides a traceable ``paper trail'' from the legal provision to its structured form. Stakeholders, such as compliance officers, can inspect the resulting code, run unit tests to confirm fidelity, and revise specific attributes if needed. This fosters greater trust in the process, as the representation is both human-auditable and machine-readable, encouraging stronger collaboration between legal and engineering experts.

Overall, the results highlight the practical feasibility of code-based representations to capture the complex structures of legal requirements. We acknowledge that our representation is not a full semantic equivalent of the legal text but a structured abstraction of it. By limiting the scope of representation to a closed class structure, we minimize ambiguity and ensure that downstream tasks (e.g., compliance checks or traceability) are grounded in testable, interpretable representations.

\subsection{Generalizability of the Python Class Structure}

The current Python class structure was derived through open coding of 13 U.S. state data breach notification laws. It primarily reflects the structural and semantic patterns found in the laws that govern personal information protection and breach notification obligations. Therefore, some degree of adaptation may be necessary when applying this structure to laws from different legal domains or jurisdictions with distinct drafting conventions. However, the following features of the class structure support its generalizability:

\begin{itemize}
    \item Domain-agnostic core classes: Classes such as Section, Statement, and Expression represent hierarchical and atomic units of legal text that are fundamental across most legal documents. These form the structural backbone of the representation and are unlikely to change drastically.
    \item Flexible semantic encoding: The Rule, Definition, Exemption, and Information classes were designed to capture commonly occurring modalities and constraints. While their current form covers obligations, permissions, prohibitions, and exceptions found in data breach laws, more specialized domains may require the addition of new classes or enrichment of existing ones (e.g., more detailed condition types).
    \item Extensible attributes and optional defaults: All class attributes are optional with default values, which ensures that code generated from laws lacking certain elements (e.g., no explicit conditions or exemptions) will still compile and remain structurally valid. This design supports partial representations and minimizes breakage when new attributes are introduced.
    \item Additive evolution: The design can be updated by introducing new classes or attributes to accommodate unseen elements without invalidating existing representations. 
\end{itemize}

\section{Threats to Validity}
\label{section:threats}

\textit{Construct Validity} refers to whether we are truly measuring what we believe we are measuring~\cite{Yin09}. We use a comprehensive test suite to verify the syntactic and semantic attributes of the code representation. While these tests capture many possible errors, passing all tests may not guarantee complete semantic equivalence with the original legal text. Deeply nested or context-dependent provisions could still go undetected if not encoded in our test cases. Furthermore, the ground truth relies on human annotations when translating legal provisions into Python class objects. Although these annotations were verified through iterative checks and unit test execution, there is a risk that errors may occur. We mitigated this by testing the comparability of the manually written code. The first author also revisited the annotations to identify and correct any inconsistencies.

\textit{Internal validity} refers to validity of analyses and conclusions drawn from the data~\cite{Yin09}. In addition, the final class structure was derived through open coding and iterative refinement. Although the schema was designed to capture core legal metadata (e.g., definitions, obligations, prohibitions, exemptions), other researchers might define or group attributes differently based on alternative theoretical frameworks or application requirements.

\textit{External validity} refers to the generalizability of results~\cite{Yin09}. We employed GPT-4o and model-specific behavior, such as tokenization, learned embeddings, and instruction tuning, affect how the model reponds to a prompt and how it handles long-range dependencies. A different model family or a later model release from the same family should be expected to yield different results, limiting exact reproducibility and requiring adjustments to the prompts. 

The experimental dataset includes 13 U.S. state data breach notification statutes, all written in English. Although these laws vary in length and drafting style, they represent only a subset of the broader legal landscape. Generalizing our results to other jurisdictions (e.g., non-English texts, international regulations) or domains (e.g., tax law, environmental regulations) may require additional tuning or domain-specific exemplars. Although we selected diverse state statutes, focusing on data breach notification requirements may limit our findings' applicability to statutes with significantly different thematic content or structural complexity (e.g., intellectual property law).

\section{Conclusion and Future Work}
\label{section:conclusion}

In this paper, we presented a novel method to automatically generate a structured, executable representation of legal text using GPT-4o. By combining textual entailment and in-context learning, our approach preserves both the structural hierarchy and rich semantic information within legal provisions. We integrated this information into a Python class structure that can be parsed, analyzed, and tested programmatically.

Evaluations on 13 US State data breach notification laws show that our approach achieves high correctness scores, with near-perfect compilation rates and strong precision and recall across most metadata attributes. The experiments further suggest that code-based representations outperform conventional text-based outputs (e.g., JSON), especially for tasks that demand explicit preservation of the original legal text, hierarchical nesting, and cross-references. Additionally, we demonstrated the approach’s ability to generalize to new, unseen statutes with minimal performance degradation.

Future work will expand this framework to other regulatory domains, and explore more sophisticated retrieval and prompting algorithms for demonstration selection. We envision using this ``executable'' legal translation for downstream analysis of compliance verification, including requirement tracing, conflict detection, and creating knowledge graphs to show how engineers interpret law in designing accountable software systems.

\section*{Data Availability Statement}
The code and data used in this study are publicly available at Zenodo\footnote{https://doi.org/10.5281/zenodo.15794182}\cite{SB25}. The repository includes source code, legal text datasets, model outputs, evaluation scripts, and instructions for reproduction.

 \section*{Acknowledgment}
This research was sponsored in part by NSF Award \#2217572.

\end{document}